\documentclass[aps,prl,twocolumn,
amsmath,showpacs]{revtex4}
\usepackage{graphicx,color,amsmath}
\begin{document}
\title{Angular distributions in three-body baryonic $B$ decays}
\author{C. Q. Geng and Y. K. Hsiao} 
\affiliation{Department of Physics,
National Tsing Hua University, Hsinchu, Taiwan 300}
\date{\today}
\begin{abstract}
We study the angular distributions of the baryon-antibaryon
low-mass enhancements in the three-body baryonic B decays of $B\to
p\bar{p}M$ ($M=K$ and $\pi$) in the framework of the perturbative
QCD. By writing the most general forms for the transition form
factors of $B\to p\bar{p}$,
we find that the angular distribution asymmetry
in $B^-\to p\bar{p}K^-$ measured by the Belle Collaboration
can be explained.
We give a quantitative description on the Dalitz plot asymmetry   
in $B^-\to p\bar{p}K^-$ shown by the BaBar Collaboration
and demonstrate that it
is equivalent to the  angular asymmetry.
In addition, we present our results on $\bar B^0\to p\bar{p}K_S$
and $B^-\to p\bar{p}\pi^-$ and we obtain that their
angular asymmetries are $-0.35\pm0.11$ and $0.45\pm0.10$,
respectively, which can be tested by the ongoing experiments at
the $B$ factories.
\end{abstract}

 \maketitle
\newpage
The three-body baryonic $B$ decays of $B\to p\bar{p}M\ (M=K^{(*)}\,,\pi)$
with a near threshold enhancement in the $p\bar{p}$ invariant mass spectrum
have been  observed by the Belle~\cite{ExptBelle,BelleAD} and BaBar~\cite{ExptBaBar} Collaborations.
In fact, the dibaryon threshold enhancement has been also observed
in other decays, such as the charmless modes
$\bar{B}^0\to\Lambda\bar{p}\pi^+$, $B^-\to\Lambda\bar{p}\gamma$
and $B^-\to\Lambda\bar{\Lambda}K^-$ \cite{Charmless,radiative} and
the charmful ones $\bar{B}^0\to p\bar{p}D^{(*)0}$, $\bar{B}^0\to
n\bar{p}D^{*+}$ and $B^-\to\Lambda^+_c\bar{p}\pi^-$ \cite{Charm}
as well as the $J/\Psi$ decays $J/\Psi\to p\bar{p}\gamma$ and
$J/\Psi\to\bar{\Lambda}pK^-$ \cite{JPhi}. 
Theoretically, the dibaryon threshold enhancements in the three-body baryonic $B$ decays were first conjectured in Ref. \cite{HouSoni}. 
Subsequently, various interpretations, including
models with a baryon-antibaryon bound state or baryonium
\cite{Baryonium}, exotic glueball states \cite{Glueball,Rosner},
fragmentation \cite{Rosner} and final
state interactions \cite{FSI} have been proposed. 
Furthermore, the  enhancements in 
$B\to {\bf B}\bar{\bf B}'M\ ({\bf
B,B'}=p,\Lambda$ and $M=D,K,\pi,\gamma)$ have been
successfully understood \cite{CK1-CK2,CK3,GengR,Geng,Geng1}
within the framework of the perturbative QCD (PQCD) based on the
QCD counting rules \cite{QCD1,QCD2} due to the power
expanded baryonic form factors.
On the other hand, the threshold enhancements in the
$J/\Psi$ decays are still in favor of some bound states
\cite{JPhi}. 

To find out the origin of the
threshold enhancements and distinguish among the above theoretical
models, various experimental studies have been performed 
\cite{BabarThesis}. 
 The Belle
Collaboration has studied the angular
distribution asymmetry~\cite{BelleAD}
in the
helicity frame for the decays of $B^-\to p\bar{p}K^-$, $\bar
B^0\to p\bar{p}K_S$ and $\bar{B}^0\to\Lambda\bar{p}\pi^+$,
while the BaBar Collaboration has measured the Dalitz plot asymmetry 
\cite{ExptBaBar} in the decay of $B^-\to p\bar{p}K^-$,
which are sensitive to the physical nature at the threshold as well
as the decay mechanism.
In particular, both data support the quark fragmentation mechanism but
disfavor the gluonic picture. 
 So far, there is no consistent understanding of the two asymmetries
and thus one cannot directly compare 
the data between the two Collaborations.

%

The angular distribution in $\bar{B}^0\to\Lambda\bar{p}\pi^+$ has
been examined in Refs. \cite{CK3,Tsai} based on the QCD counting
rules and the result
is consistent with the data
\cite{BelleAD}. 
However, such an approach faces a difficulty
\cite{Tsai,HYreview1} to understand the shapes for the
modes of $B\to p\bar{p}K$ \cite{BelleAD} due to the lack of
understanding of the baryonic transition form factors in $B\to
p\bar{p}$.
Moreover, quantitative descriptions of the Dalitz plot asymmetries are not yet available besides qualitative quark fragmentation mechanism \cite{Rosner}. 

In this letter, we
will concentrate on the three-body charmless baryonic decays of
$B\to p\bar{p}M\ (M=K,\pi)$. We will first write down the most
general form of the $B\to {\bf B\bar B'}$ transition matrix
element and then fix the unknown form factors by the experimental
data. 
In our study, we will use the QCD counting rules as well as
the $SU(3)$ flavor symmetry to get and relate the behaviors of the
form factors. 
We will demonstrate that when these form factors are
constructed, the data of the angular distribution \cite{BelleAD}
 and Dalitz plot \cite{ExptBaBar} asymmetries
  in $B^-\to p\bar{p}K^-$  can be understood. 
  We will show that the two asymmetries are equivalent by
  describing the same physics.
  Furthermore, we will
 extend our investigation to the decays of $\bar B^0\to
p\bar{p}K_S$ and $B^-\to p\bar{p}\pi^-$.

 From the effective Hamiltonian at quark level
\cite{Hamiltonian},
the decay amplitude of $B^-\to p\bar p K^-$ is
separated into two parts, given by
\begin{eqnarray}
\label{amp}
{\cal A}(B^-\to p\bar p K^-)=
{\cal C}(B^-\to p\bar p K^-)+{\cal
T}(B^-\to p\bar p K^-)\,,
\nonumber\\
{\cal C}(B^-\to p\bar p K^-)=\frac{G_F}{\sqrt 2}\bigg\{
V_{ub}V_{us}^* a_2\langle p\bar p|(\bar u u)_{V-A}|0\rangle-\ \ \
\nonumber\\
V_{tb}V_{ts}^*\bigg[a_3\langle p\bar p|(\bar u u+\bar d d)_{V-A}|0\rangle
+a_5\langle p\bar p|(\bar u u+\bar d d)_{V+A}|0\rangle
\nonumber\\
+\frac{a_9}{2}\langle p\bar p|(2\bar u u-\bar d
d)_{V-A}|0\rangle\bigg] 
\bigg\}\langle K^-|(\bar s b)_{V-A}|B^-\rangle\;,\ 
\nonumber
\\
{\cal T}(B^-\to p\bar p K^-)=\frac{G_F}{\sqrt 2}\bigg\{
(V_{ub}V_{us}^* a_1-V_{tb}V_{ts}^*a_4)
\ \ \ \ \ \ \ \ \ \ 
\nonumber\\
\langle K^-|(\bar s u)_{V-A}|0\rangle \langle p\bar p|(\bar u
b)_{V-A}|B^-\rangle
+V_{tb}V_{ts}^*2a_6\ \ \ \ \ \
\nonumber\\
\langle K^-|(\bar s u)_{S+P}|0\rangle 
\langle
p\bar p|(\bar u b)_{S-P}|B^-\rangle \bigg\}\;,\ \ \ \
\end{eqnarray}
where  $G_F$ is the Fermi constant, $V_{q_iq_j}$ are the CKM
matrix elements,
$(\bar{q}_iq_j)_{V-A}=\bar{q}_i\gamma_\mu(1-\gamma_5)q_j$,
$(\bar{q}_iq_j)_{S\pm P }=\bar{q}_i(1\pm\gamma_5)q_j$ and
$a_i=c^{eff}_i+c^{eff}_{i\pm 1}/N_c$ for $i=$odd (even) in terms
of the effective Wilson coefficients $c^{eff}_i$, defined in Refs.
\cite{Hamiltonian,ali}, and the color number $N_c$. In Eq.
(\ref{amp}), we have assumed the factorization approximation. As
seen in Eq. (\ref{amp}), the current part of the amplitude
involves time-like baryonic form factors from $\langle p\bar
p|(\bar{q}q)_{V,A}|0\rangle$ and $B\to K$ transition form factors,
which can be referred in Refs. \cite{Geng1} and \cite{BtoKff},
respectively. On the other hand, for the transition part of the
amplitude, we need to evaluate the transition matrix elements of
$B\to p\bar{p}$. Based on Lorentz invariance, the most general
forms for the $B\to {\bf B\bar B'}$ transition matrix elements due
to scalar, pseudoscalar, vector and axial-vector currents  are
given by
\begin{eqnarray}
&&\langle {\bf B}{\bf\bar B'}|S^b|B\rangle
=i\bar u(p_{\bf B})[f_A \not{\!p}+f_P]\gamma_5v(p_{\bf \bar B'})\;,
\nonumber\\
&&\langle {\bf B}{\bf\bar B'}|P^b|B\rangle
= i\bar u(p_{\bf B})[f_V\not{\!p}+f_S]v(p_{\bf \bar B'})\;,
\nonumber\\
&& \langle {\bf B}{\bf\bar B'}|V^b_{\mu}|B\rangle=
i\bar u(p_{\bf B})[g_1\gamma_{\mu}+g_2i\sigma_{\mu\nu}p^\nu+
g_3p_{\mu}
\nonumber\\
&& \ \ +g_4(p_{\bf\bar B'}+p_{\bf B})_\mu +g_5(p_{\bf\bar B'}-p_{\bf B})_\mu]\gamma_5v(p_{\bf \bar B'})\;,\nonumber\\
&&\langle {\bf B}{\bf\bar B'}|A^b_{\mu}|B\rangle= i\bar u(p_{\bf B})[f_1\gamma_{\mu}+f_2i\sigma_{\mu\nu}p^\nu+f_3p_{\mu}\nonumber\\
&&\ \ +f_4(p_{\bf\bar B'}+p_{\bf B})_\mu +f_5(p_{\bf\bar B'}-p_{\bf B})_\mu]v(p_{\bf \bar B'})\;,
\label{BtoBB}
\end{eqnarray}
respectively, where $S^b=\bar q b$, $P^b=\bar q\gamma_5 b$,
$V^b_\mu=\bar q\gamma_\mu b$ and $A^b_\mu=\bar q\gamma_\mu\gamma_5
b$ with $q=u,d$ and $s$, $p=p_B-p_{\bf B}-p_{\bf\bar B'}$ is the
emitted four-momentum, and $g_i$ and $f_i$ are the the form
factors to be determined. Here the parity conservation in strong
interactions is used. We note that the forms for the scalar and
pseudoscalar currents in Eq. (\ref{BtoBB}) have been studied
($e.g.$, see Ref. \cite{CK1-CK2}), but those involving the vector and
axial-vector currents in Eq. (\ref{BtoBB}) have not been given in
the literature previously.

At the scale of $m_b\simeq 4$ GeV, the $t\equiv (p_{\bf
B}+p_{\bf\bar B'})^2$ dependences of the form factors in Eq.
 (\ref{BtoBB}) can be parametrized according
to the power counting rules of the PQCD \cite{Brodsky1,Brodsky2}
based on the hard gluons needed in the process. For example, in
the transition of $B\to {\bf B \bar B'}$, three hard gluons are needed
to produce ${\bf B \bar B'}$, in which two of them create the valence
quark pairs and the third one is responsible for kicking the
spectator quark in $B$ \cite{Tsai}. Since each of the gluons has a
propagator $\sim 1/q^2$ with $q^2$ proportional to the momentum of
the ${\bf B\bar B'}$ pair, all of the form factors $f_i$ and $g_i$
have to fall off as $1/t^3$, in which the propagator can be
realized to contain a zero-mass pole inducing threshold
enhancement \cite{HouSoni}. Explicitly, we have
\begin{eqnarray}
g_i=\frac{C_{g_i}}{t^3}\,,\;f_i=\frac{C_{f_i}}{t^3}\,,
\label{NewFF}
\end{eqnarray}
where $C_{g_i}$ and $C_{f_i}$ are new sets of form factors to be determined.

From equation of motion, we can relate the form factors in Eq.
 (\ref{BtoBB}) and we obtain
\begin{eqnarray}
\label{BtoBBSP2}
m_bf_A=g_1\;,\;m_bf_P=m_B [g_4E_M+g_5(E_{\bf \bar B'}-E_{\bf B})]\;,
\nonumber\\
m_bf_V=f_1\,\;,\;m_bf_S=m_B [f_4E_M+f_5(E_{\bf \bar B'}-E_{\bf B})]\;,
\end{eqnarray}
where $E_M$, $E_{\bf \bar B'}$ and $E_{\bf B}$ are the energies of
the $M$ meson, ${\bf \bar B'}$ and ${\bf B}$, respectively.
In Eq. (\ref{BtoBBSP2}),
$g_2$ and $f_2$ disappear, while $g_3$ and $f_3$ are neglected since the corresponding terms are proportional to
$m_{\bf M}^2$ which is small comparing to $m_BE_M$ and $m_B(E_{\bf
\bar B'}-E_{\bf B})$.
 The form factors in Eq.
(\ref{NewFF}) 
 can be related by
the spin $SU(2)$ and flavor $SU(3)$ symmetries.
In Table \ref{T1}, we show the relations for the form
factors in $\langle p\bar p|(\bar u b)_{V,A}|B^{-,0}\rangle$.
\begin{table}[h!]
\caption{ Relations between the $B\to p \bar p$ transition form factors with $i=2,\cdots,5$.}\label{fff}
\label{T1}
\begin{tabular}{|c||c|c|c|c|c|}
\hline
{Form Factor}&$C_{g_1}$&$C_{f_1}$&
$C_{g_i}=-C_{g_i}$
\\\hline
$\langle p\bar p|(\bar u b)_{V,A}|B^-\rangle$ 
&$\frac{5}{3}N_{||}-\frac{1}{3}N_{\overline{||}}$
&$\frac{5}{3}M_{||}+\frac{1}{3}M_{\overline{||}}$
&$\;\;\;\frac{4}{3}M^i_{||}$ 
\\
$\langle p\bar p|(\bar d b)_{V,A}|\bar B^0\rangle$ 
&$\frac{1}{3}N_{||}-\frac{2}{3}N_{\overline{||}}$
&$\frac{1}{3}M_{||}+\frac{2}{3}M_{\overline{||}}$
&$-\frac{1}{3}M^i_{||}$ 
\\
\hline
\end{tabular}
\end{table}
%
%

The decay width $\Gamma$ of $B^-\to p\bar{p}K^-$
is given by \cite{eqofag}
\begin{eqnarray}\label{Gamma}
\Gamma=\int^{+1}_{-1}\int^{(m_B-m_K)^2}_{4m_p^2}\frac{\beta_p\lambda^{1/2}_t}{(8\pi
m_B)^3}|\bar {\cal A}|^2\;dt\;d\text{cos}\theta\;,
\end{eqnarray}
where  $\beta_p=(1-4m_p^2/t)^{1/2}$,
$\lambda_t=m_B^4+m_K^4+t^2-2m_K^2 t-2m_B^2 t-2m_K^2 m_B^2$,
 $\theta$ is the angle between the three-momenta
 of the $K$ meson
and 
the proton in the dibaryon rest
frame, and $|\bar{\cal  A}|^2$ is the squared amplitude of Eq.
(\ref{amp})  by summing over all spins.
 Note that $4m_{B}E_{p(\bar p)}=m_B^2+t-m_p^2\pm
\beta_p\lambda^{1/2}_t\cos\theta$.
 From Eq. (\ref{Gamma}), we can study the partial decay width
$d\Gamma/d\cos\theta$ as a function of $\cos\theta$, $i.e.$, the
angular distribution. We may also define the 
angular asymmetry by
\begin{eqnarray}\label{AFB}
A_{\theta}\equiv\frac{\int^{+1}_0\frac{d\Gamma}{d\cos\theta}d\cos\theta
-\int^0_{-1}\frac{d\Gamma}{d\cos\theta}
d\cos\theta}{\int^{+1}_0\frac{d\Gamma}{d\cos\theta}
d\cos\theta+\int^0_{-1}\frac{d\Gamma}{d\cos\theta}
d\cos\theta}\;,
\end{eqnarray}
which is equal to $(N_+-N_-)/(N_++N_-)$, where $N_\pm$ are the
events with $\cos\theta>0$ and $\cos\theta<0$, respectively. The
angular asymmetry in $B^-\to p\bar{p}K^-$ has been
measured by the Belle Collaboration \cite{BelleAD} 
to be
\begin{eqnarray}\label{AFBK}
A_{\theta}(B^-\to p\bar p K^-)&=&0.59^{+0.08}_{-0.07}\;,
\end{eqnarray}
implying that the protons are emitted  along the $K^-$ direction
most of the time \cite{BelleAD} in the $p\bar{p}$ rest frame,
which seems to be unexpected from the previous $B$ studies in the
PQCD picture \cite{HYreview1}.

In general, when a decay mode mixing with vector
(axial-vector) and scalar (pseudoscalar) currents, it makes the
partial decay width as a form of
\begin{eqnarray}\label{dGcos}
\frac{d\Gamma}{d\cos\theta}=a\,\cos^2\theta+b\,\cos\theta+c.
\end{eqnarray}
For $\bar B^0 \to \Lambda\bar p\pi^+$,
 one gets that $c>-a>0$ with $b\simeq 0$ \cite{CK3}.
Therefore, its angular
distribution 
as a function of
$\cos\theta$ is figured as a parabolic curve opening downward
\cite{Tsai,HYreview1}, which is consistent with the data
\cite{BelleAD}. On the other hand, the angular distribution of
$B^-\to p\bar p K^-$ \cite{BelleAD} is gradually bent up as
$\text{cos}\theta=-1$ shifts to $+1$ (see Fig 3a in Ref.
\cite{BelleAD}), leading to an asymmetric $A_{\theta}$ in Eq.
(\ref{AFBK}), which is unexpected since the decay is dominated
by  $V\cdot V$ and $A\cdot A$ contributions
\cite{CK1-CK2,Geng}.
Clearly, the data for $B^-\to p\bar p K^-$ indicates
a non-neglected $\cos\theta$ term which is comparable with the
$\cos^2\theta$ one in Eq. (\ref{dGcos}), $i.e.$, $a\simeq b>0$. To
find out a large $\cos\theta$ term, it is important to note that
the energy difference of the proton pair $E_{\bar p}-E_p$ is
proportional to $\cos\theta$ and related to $g_5$ and $f_5$ as
seen from Eq. (\ref{BtoBBSP2}),
which could provide new source of the angular dependence.
We now summarize all possible  $\cos\theta$ and $\cos^2\theta$ terms
in $B^-\to p\bar p K^-$ as follows:
\begin{eqnarray}
V_1 \cdot V_5\;,\;V_4 \cdot V_5\;,\;A_1 \cdot A_4\;,\;A_4 \cdot A_5\;
&\propto&\;\cos\theta\,,
\nonumber\\
V_{1(5)} \cdot V_{1(5)}\;,A_{1(5)} \cdot A_{1(5)}\;,\;A_1 \cdot A_5\;,\;&\propto&\;\cos^2\theta\,,
\end{eqnarray}
where we have denoted the terms corresponding to $g_i$ and $f_i$ in
Eq. (\ref{BtoBB}) as $V_i$ and $A_i$, respectively.
The squared amplitude in Eq. (\ref{Gamma}) is reduced to be
\begin{eqnarray}
{\it |\bar {\cal A}|^2}&=&\bigg(\frac{G_F}{\sqrt 2}f_Km_B^3
\alpha_K\bigg)^2 (\rho_0 +\rho_\theta\cos\theta
+\rho_{\theta^2}\cos^2\theta ),
\nonumber\\
\rho_0&=&
\frac{g_1^2}{2m_B^2}(1-\hat{t})^2
+4{g_1\over m_B} g_4\hat{E}_{K}\hat{m}_p (1-\hat{t})
+2g_4^2 \hat{E}_K^2 \hat{t}
\nonumber\\
&&+\bigg(\frac{\beta_K}{\alpha_K}\bigg)^2
\bigg[\frac{f_1^2}{2m_B^2}(1-\hat{t})^2+2\hat{E}_K^2 f_4^2(\hat{t}-4\hat{m}_p^2)\bigg]\;,\nonumber\\
\rho_\theta&=&2\beta_p \hat{\lambda}^{1/2}_t
\bigg\{{g_1 g_5\over m_B} \hat{m}_p (1-\hat{t})+g_4 g_5 \hat{E}_K \hat{t}
+\nonumber\\
&&\bigg(\frac{\beta_K}{\alpha_K}\bigg)^2 
\hat{E}_K\bigg[2 {f_1 f_4\over m_B}  \hat{m}_p
+ f_4 f_5 (\hat{t}-4 \hat{m}_p^2)\bigg]\bigg\}\;,\nonumber\\
\label{squaredA2-3}
\rho_{\theta^2}&=&\frac{1}{2}\beta_p^2\hat{\lambda}_t\bigg\{
-{g_1^2\over m_B^2}+g_5^2\hat{t}+\bigg(\frac{\beta_K}{\alpha_K}\bigg)^2\nonumber\\
&&
\bigg[-{f_1^2\over m_B^2}+4{f_1 f_5\over m_B}\hat{m}_p+f_5^2(\hat{t}-4
\hat{m}_p^2)\bigg]\bigg\}\;,
\end{eqnarray}
with $(\hat{t},\hat{\lambda}^{1/2}_t)=(t,\lambda^{1/2}_t)/m_B^2$, 
$(\hat{m}_p,\hat{E}_K)=(m_p,E_K)/m_B$ and
\begin{eqnarray}\label{squaredA3}
\alpha_K(\beta_K)=V_{ub}V_{us}^* a_1-V_{tb}V_{ts}^*\bigg[a_4\pm \frac{2a_6m_K^2}{m_b(m_s+m_u)}\bigg].
\end{eqnarray}
Here, to simplify our formula  we have ignored the current contribution of
${\cal C}(B^-\to p\bar p K^-)$ in Eq. (\ref{amp}), 
estimated around $1\%$ of the measured one by using the values of
the time-like baryonic form factors in \cite{Geng1}.
 However, all terms are kept
in our numerical calculation.

\begin{figure}[t!]
\centering
\includegraphics[width=1.6in]{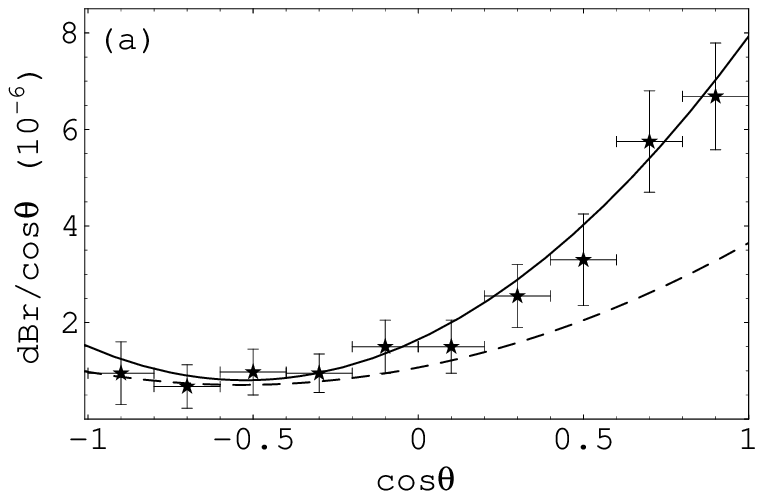}
\includegraphics[width=1.6in]{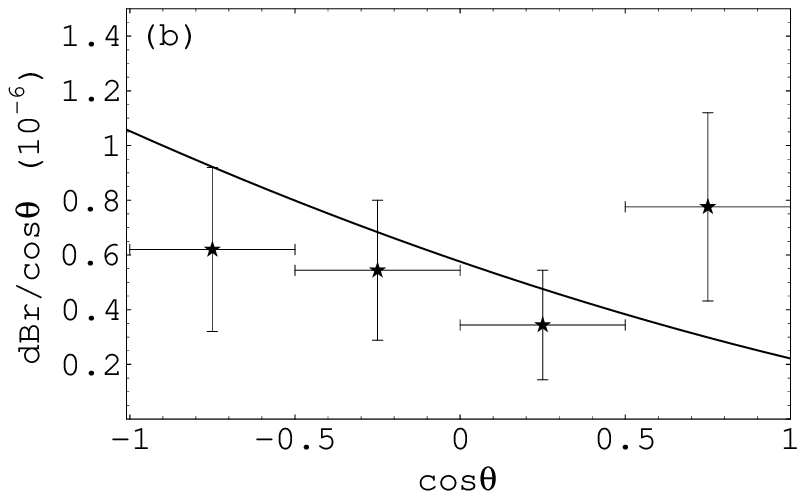}
\caption{\label{fig} Branching fraction vs. $\cos\theta$ in the
$p\bar{p}$ rest frame for (a) $B^-\to p\bar p K^-(\pi^-)$ with
solid (dash) curve and (b) $\bar B^0 \to p\bar p K_S$,
where the data
points (stars) in (a) and (b) are from Ref. \cite{BelleAD}.
 }
\end{figure}

In our numerical analysis, we take the 10 data points of
$dBr/d\cos\theta$ in $B^-\to p\bar p K^-$ from Fig. 3a
in Ref. \cite{BelleAD} (see also Fig. \ref{fig}) and measured decay branching ratios \cite{BelleAD} of
$(5.74\pm 0.61)$
and
$(1.20\pm 0.35)\times 10^{-6}$
for $B^-\to p\bar p K^-$ and $\bar B^0\to p\bar p K_S$, respectively,
as input values, along with the CKM matrix elements referred in
Ref. \cite{pdg},
$(a_1,a_4,a_6)=(1.05,-0.0441-0.0072i,-0.0609-0.0072i)$
\cite{Hamiltonian}, $m_u(m_b)=3.2$ MeV, $m_s(m_b)=90$ MeV
\cite{Hamiltonian} and $m_b(m_b)=4.19\pm 0.05$ GeV \cite{mbmass}.
We note that the ratio of $(\beta_K/\alpha_K)^2$ is around $15\%$
which suppresses $f_{V,S}$ (or $f_{1,4,5}$) terms in Eqs.
 (\ref{BtoBB}) and (\ref{BtoBBSP2}). Although
there is no surprise that $V_1\cdot V_1$ and $A_1\cdot A_1$ create
$\cos^2\theta$ with an expected minus sign, the dominant
contribution arises from the $g_5^2$ term with $t$ around $4-25$
GeV$^2$ and as a consequence, 
$\rho_\theta$ and $\rho_{\theta^2}$ can be both
positive with the same size and  the condition of $a\simeq b>0$ is
fulfilled. By performing the fitting, $N_{||,\bar{||}}$
and $M_{||}^{4,5}$ in Table \ref{T1}
are determined to be
\begin{eqnarray}
N_{||}=127.1\pm 26.6\, GeV^5,\,N_{\overline{||}}=-200.9\pm 51.9\,GeV^5,\ \nonumber\\
M_{||}^4=-25.0\pm 15.4\,GeV^4,\,M_{||}^5=227.3\pm 22.0\,GeV^4,\  \
\end{eqnarray}
As seen in Fig. \ref{fig}a, our result (solid curve) for
$dBr(B^-\to p\bar p K^-)/d\cos\theta$ as a function of
$\cos\theta$ explains the data \cite{BelleAD} well
and it is in agreement with the Belle data of $A_{\theta}$ in Eq.
(\ref{AFBK}).
 For $\bar B^0\to p\bar p K_S$, the
angular distribution is shown in Fig. \ref{fig}b, which is also
consistent with the data except the one point close to
$\cos\theta=0.8$. Clearly, more 
data at the ongoing
$B$ factories are needed. In addition,
we predict 
\begin{eqnarray}\label{pred}
A_{\theta}(B^-\to p\bar p\pi^-)&=&0.45\pm 0.11\;,\nonumber\\
A_{\theta}(\bar B^0\to p\bar p K_S)&=&-0.35\pm 0.10\;.
\end{eqnarray} 
The result in Eq. (\ref{pred}) for $A_{\theta}(B^-\to p\bar p\pi^-)$ is
anticipated  since the decay is similar to $B^-\to p\bar p K^-$
although it is dominated by the tree-diagram. It is interesting
to point out that 
the minus sign for $A_{\theta}(\bar B^0\to
p\bar p K_S)$ in Eq. (\ref{pred})
is different from the expectation of the fragmentation mechanism 
\cite{Rosner}.
 We note that from
our fitted form factors we  obtain
\begin{eqnarray}
\label{pred1}
Br(B^-\to p\bar p \pi^-)&=&(3.0\pm 0.4)\times 10^{-6}\,,
\end{eqnarray}
which supports the Belle measurement of $(3.06^{+0.73}_{-0.62}\pm
0.37)\times 10^{-6}$ \cite{ExptBelle}, but is higher than the
Babar data of $(1.24\pm 0.32\pm 0.10)\times 10^{-6}$
\cite{BabarThesis}. We look forward to having the future
experiments at Belle and BaBar to check our predictions
in Eqs. (\ref{pred}) and (\ref{pred1}).

\begin{figure}[t!]
\centering
\includegraphics[width=1.67in]{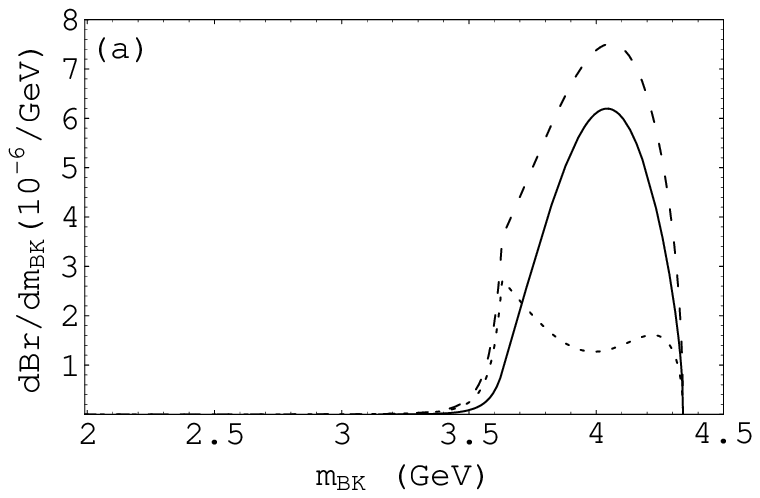}
\includegraphics[width=1.65in]{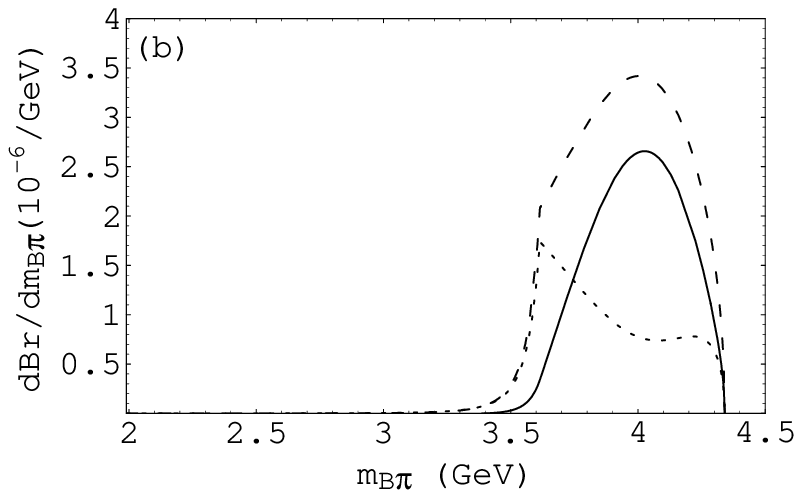}
\caption{\label{fig2} 
Branching fraction of $Br(B^-\to p\bar p M^-)/dm_{{\bf B}M}$
with (a) $M=K$ and (b) $M=\pi$ 
as functions of $m_{{\bf B}M}$ (${\bf B}=\bar p,p$) 
with the dash, dot and solid curves representing
(i) $m_{\bar{p}M^-}>m_{pM^-}$, (ii) $m_{\bar{p}M^-}<m_{pM^-}$ and
(iii) difference between (i) and (ii), respectively.
 }
\end{figure}

It is known that
the asymmetric Dalitz plot for $B^-\to p\bar p K^-$ reported by
the BaBar Collaboration \cite{ExptBaBar,BabarThesis}
supports the fragmentation mechanism \cite{Rosner}
but disfavors the gluonic resonance state \cite{Glueball,Rosner}
as well as other intermediate state in the pole model \cite{HYpole}.
We now examine if we can give a quantitative description on the 
Dalitz plot in the PQCD approach. To do this, we define the
Dalitz plot asymmetry for $B\to {\bf B \bar B'}M$ as 
\begin{eqnarray}
{\it A}_{DP}=\frac{\Gamma(m_{\bf \bar B' M})_>-\Gamma(m_{\bf B
M})_>}{\Gamma(m_{\bf \bar B' M})_>+\Gamma(m_{\bf B M})_>}\;,
\end{eqnarray}
where $\Gamma(m_{\bf B M})_>$ $[\Gamma(m_{\bf \bar B' M})_>]$ 
denotes the decay width for the range of $m_{\bf B M}>m_{\bf \bar B'M}$
($m_{\bf \bar B'M}>m_{\bf B M}$) divided by the line of $m_{\bf B
M}=m_{\bf \bar B'M}$ in the Dalitz plot. It is easy to see
that for $B^-\to p\bar p K^-$, $i.e.$, ${\bf B}={\bf B'}$,
the Dalitz plot asymmetry is identical to the angular asymmetry
in Eq. (\ref{AFB})
due to the fact that these two
asymmetries arise from the same source of $\rho_\theta$ in
Eq. (\ref{squaredA2-3}) with the relations of
$\text{cos}\theta=\beta_p^{-1}\lambda^{-1/2}(m_{\bar p
K^-}^2-m_{pK^-}^2)$ and $t=m^{2}_{B}+2m^{2}_{p}+m_{K}^{2}-(m^{2}_{\bar{p}K^{-}}+m^{2}_{pK^{-}})$.
Explicitly, in Fig. \ref{fig2}a, we show the decay
branching fraction of $B^-\to p\bar p K^-$ as functions of $m_{\bar{p}K^{-},pK^{-}}$ 
with the dash, dot and solid curves representing
(i) $m_{\bar{p}K^-}>m_{pK^-}$, (ii) $m_{\bar{p}K^-}<m_{pK^-}$ and
(iii) difference between (i) and (ii), respectively.
 It is interesting to note that, as seen from the figure,
the Dalitz plot asymmetry peaked around 4 GeV is exactly the
same as the data in Ref. \cite{ExptBaBar}. However, our prediction
for $B^-\to p\bar p \pi^-$ shown in Fig. \ref{fig2}b is different from
the BaBar unpublished result in \cite{BabarThesis} like
the decay branching fraction. Clearly, more precise data for the
Dalitz plot distribution on the $\pi$ mode are needed.

Finally, we remark that the 
form factors in
Eq. (\ref{BtoBB}) determined by the angular distribution in
$B^-\to p\bar p K^-$ can be used to examine other 
experimental measured
three-body baryonic
 $B$ decays with a vector meson in the final state, such as
$B\to p\bar p K^{*}$
and $B^-\to
\Lambda \bar p J/\Psi$ \cite{ExptBelle,Charm}, which have not been 
theoretically explored yet. Furthermore, 
 direct CP violation in $B^-\to p\bar p K^{(*)-}$
can be also investigated.
Moreover, our study on the
angular distribution can be extend to the above modes as well as 
the decay of 
 $B^-\to \Lambda\bar p \gamma$, 
which has only been discussed in the pole model
\cite{HYradi2}.

We thank Drs. H.Y. Cheng, M. Graham and M.Z. Wang for useful discussions.
This work is supported by the National Science Council of Republic
of China under the contract \#s NSC-94-2112-M-007-(004,005).

\end{document}